\documentclass[12pt]{article}
\usepackage{graphicx}

\def \bea{\begin{eqnarray}}
\def \beq{\begin{equation}}

\def \eea{\end{eqnarray}}
\def \eeq{\end{equation}}

\begin{document}
\rightline{CLNS 04/1873}
\rightline{hep-ph/0404245}
\bigskip

\centerline{\bf VECTOR MESON PAIR PRODUCTION IN}
\centerline{\bf TWO-PHOTON COLLISIONS NEAR THRESHOLD
\footnote{Submitted to Phys.\ Rev.\ D.}}
\bigskip
 
\centerline{Jonathan L. Rosner~\footnote{rosner@hep.uchicago.edu.  On leave
from Enrico Fermi Institute and Department of Physics,
University of Chicago, 5640 S. Ellis Avenue, Chicago, IL 60637}}
\centerline{\it Laboratory of Elementary Particle Physics}
\centerline{\it Cornell University, Ithaca, NY 14850}
 
\begin{quote}
The cross sections for several $\gamma \gamma \to VV$ processes exhibit strong
enhancements near threshold, where $V$ denotes a vector meson.  The
pattern of these enhancements is not well understood; for example, $\gamma
\gamma \to \rho^0 \rho^0$ shows an appreciable peak, while $\gamma \gamma \to
\rho^+ \rho^-$ does not.  Some possible mechanisms for this behavior are
discussed.  Tests are proposed involving production of systems containing
heavier quarks, e.g., through the reaction $\gamma \gamma \to J/\psi \rho^0$.
The importance of modeling $\ell^+ \ell^- \pi^+ \pi^-$ angular distributions in
decays of threshold $J/\psi \rho^0$ enhancements is illustrated
by comparing the expected distributions for S-wave decays of scalar particles
and P-wave decays of pseudoscalar particles.
\end{quote}

\noindent
PACS Categories: 12.39.Mk, 13.60.Le, 13.66.Bc, 14.40.Cs

\bigskip

\centerline{\bf I.  INTRODUCTION}
\bigskip

The properties of meson and baryon resonances can be strongly influenced not
only by their intrinsic quark content, but also by the final states to which
they couple.  Recently several resonances have been reported for which these 
effects may be important.  These include $J^P = 0^+$ and $1^+$ charmed-strange
mesons several tens of MeV below $DK$ and $D^* K$ threshold, respectively
\cite{Aubert:2003fg, Besson:2003cp,Abe:2003jk} and a charmonium resonance
$X(3872)$ nearly degenerate with the $D^0 \bar D^{*0}$ threshold
\cite{Choi:2003ue,Acosta:2003zx}.  Such effects may also be important for the
reactions $\gamma \gamma \to V V$, where $V$ is a vector meson.  In the
present paper I discuss some proposals to account for the curious pattern of
threshold enhancements in these processes, note the availability of the $J/\psi
\rho$ channel, and calculate angular distributions for two cases of spinless
particle decay to $J/\psi \rho$.

The cross section for $\gamma \gamma \to \rho^0 \rho^0$ is strongly enhanced
near threshold \cite{Brandelik:1980jv,Albrecht:1990cr}, while $\gamma \gamma
\to \rho^+ \rho^-$ is not \cite{Albrecht:1991sr}.  This difference appears to
occur mainly for real or nearly-real photons; when one photon is highly
virtual the cross sections for $\gamma \gamma^* \to \rho^0 \rho^0$
\cite{Achard:2003qa} and $\gamma \gamma^* \to \rho^+ \rho^-$ \cite{Achard} are
much more similar.  It will be interesting to see whether a recent calculation
of $\gamma \gamma^* \to \rho^0 \rho^0$ \cite{Anikin:2003fr} can reproduce the
result for $\gamma \gamma^* \to \rho^+ \rho^-$.

Other $\gamma \gamma \to V V$ processes exhibit patterns \cite{Albrecht:1996gr}
not all of which can be understood from a single standpoint.  A review of
$\gamma \gamma$ interactions up to 2001 may be found in Ref.\
\cite{Whalley:2001mk}.  Additional relevant experimental data
comes from a recent search for the $X(3872)$ in photon-photon collisions, in
which a sample of $J/\psi \pi^+ \pi^-$ events has been studied \cite{Zweber},
and which is the main motivation for the present study.

Photon-photon collisions share features with hadron-hadron collisions, but have
properties making them easier to interpret.  In hadronic processes at low
transverse momentum one can regard photons as superpositions of vector mesons
$\rho + \omega + \phi$ with relative couplings 3:1:$\sqrt{2}$ \cite{Leith:zd}.
SU(3) symmetry is broken through a lower total cross section of $\phi$ mesons
with non-strange hadrons. A similar approach describes $J/\psi$ photoproduction
through a suppressed $J/\psi$--nucleon cross section \cite{Carlson:1974ev}.

This paper is organized as follows.  Section II reviews the processes $\gamma
\gamma \to VV$.  I discuss some possible mechanisms for the threshold
enhancement of such processes in Section III.  Section IV is devoted to
information available from the process $\gamma \gamma \to J/\psi \rho^0$, while
Section V concludes.
\bigskip

\centerline{\bf II.  DATA ON $\gamma \gamma \to VV$}
\bigskip

The cross sections for $\gamma \gamma \to V V$ generally peak not far above
threshold.  The greatest enhancement occurs for $\rho^0 \rho^0$, for which
$\sigma(\gamma \gamma \to \rho^0 \rho^0)$ rises to a maximum of $\sim (57 \pm
6)$ nb in the photon-photon center-of-mass energy range $W = 1.5$ to 1.6 GeV
\cite{Albrecht:1990cr} and in the state with $(J=2,J_z = \pm 2)$.  No such
enhancement is seen in $\gamma \gamma \to \rho^+ \rho^-$.  The cross section
for the corresponding partial wave shows a dip in this region and never exceeds
$\sim (12 \pm 4)$ nb \cite{Albrecht:1991sr}.  The peak cross sections
$\sigma_{\rm pk}$ and the energy ranges $W$ in which they occur
\cite{ARGUS} are summarized in Table I \cite{Albrecht:1990cr,Albrecht:1991sr,%
Kriznic:1994me,Albrecht:1995nd,Albrecht:1994hg,Albrecht:1999zu}.

\begin{table}
\caption{Energy ranges and peak cross sections for $\gamma \gamma
\to V V$.
\label{tab:pks}}
\begin{center}
\begin{tabular}{l c c c } \hline \hline
$VV$ & $W$ (GeV) & $\sigma_{\rm pk}$ (nb) & Reference \\ \hline
$\rho^0 \rho^0$ & 1.5--1.6 & $57 \pm 6^a$ & \cite{Albrecht:1990cr} \\
$\rho^+ \rho^-$ & $\sim 1.3$ & $12 \pm 4^a$ & \cite{Albrecht:1991sr} \\
$\rho^0 \omega$ & 1.5--1.7 & $17.3 \pm 3.5$ & \cite{Kriznic:1994me} \\
$\omega \omega$ & 1.6--1.8 & $4.3 \pm 1.5$  & \cite{Albrecht:1995nd} \\
$\rho^0 \phi$   & 1.75--2.00 & $2.2 \pm 1.1$ & \cite{Albrecht:1994hg} \\
$\omega \phi$   & 1.9--2.3 & $1.65 \pm 0.86$ & \cite{Albrecht:1994hg} \\
$K^{*+}K^{*-}$  & 2.00--2.25 & $36.20 \pm 6.25$ & \cite{Albrecht:1999zu} \\
$K^{*0}\bar K^{*0}$ & 1.75--2.00 & $5.97 \pm 0.78$ & \cite{Albrecht:1999zu} \\
\hline \hline
\end{tabular}
\end{center}
\leftline{$^a$ In partial wave $J=2,J_z = \pm 2$.}
\end{table}

An interesting feature of several of the threshold bumps is that in cases
in which a partial-wave analysis is possible (such as $\gamma \gamma \to
\rho^0 \rho^0$ and $\gamma \gamma \to K^{*0}\bar K^{*0}$) they occur in the
$(J=2, J_z = \pm 2)$ state, with little activity in $J=0$ or $J=2, J_z = 0$.

\bigskip
\centerline{\bf III.  MECHANISMS FOR THRESHOLD ENHANCEMENTS}
\bigskip

The preponderance of $\rho^0 \rho^0$ over $\rho^+ \rho^-$ near threshold could
be due \cite{Achasov:1981kh,Li:1981ez} to a superposition of resonances with
isospins zero and two near $\rho \rho$ threshold, coupling much more strongly
to $\rho^0 \rho^0$ than to $\rho^+ \rho^-$.  If there really is a $\rho \rho$
resonance near threshold with $I=2$, there should exist states around 1600 MeV
decaying to $\rho^\pm \rho^\pm$.  While no such states have been seen so far, 
the relevant final state $\pi^\pm \pi^\pm \pi^0 \pi^0$ has not been carefully 
examined in a detector with good sensitivity to both charged and neutral 
particles.  Such detectors as CLEO, BaBar, and Belle, constructed to
investigate $e^+ e^-$ collisions at energies suitable for studying the decays
of $B$ mesons, would be excellent places to pursue such searches
\cite{Rosner:2003ia}.

It should be possible to construct a set of resonances sufficient to reproduce
the pattern of Table \ref{tab:pks}.  Some difficulty in this regard was pointed
out in Ref.\ \cite{Achasov:1989ky}.  A resonance which decays to $K^{*+}
K^{*-}$ should, in principle, also be able to decay to $\rho^0 \phi$ unless it
has isospin zero.  But then it should have the same branching ratio to $K^{*0}
\bar K^{*0}$ as to $K^{*+} K^{*-}$, which is certainly not suggested by the
pattern of peak cross sections.

Another view of the $\rho^0 \rho^0$ enhancement \cite{Bajc:1996ew}
is that each photon produces a $\rho^0$ and these vector mesons then interact
with one another through the repeated exchange of an $I=0$ meson ``$\sigma$'',
leading to an effective potential between the vector mesons.  This mechanism
is illustrated in Fig.\ 1(a).  It does not contribute to $\gamma \gamma \to
\rho^+ \rho^-$, accounting for the suppression of that process.  But it also
does not contribute to $\gamma \gamma \to K^* \bar K^*$, a shortcoming which
is particularly acute for the charged pair.

\begin{figure}
\begin{center}
\includegraphics[height=6.5in]{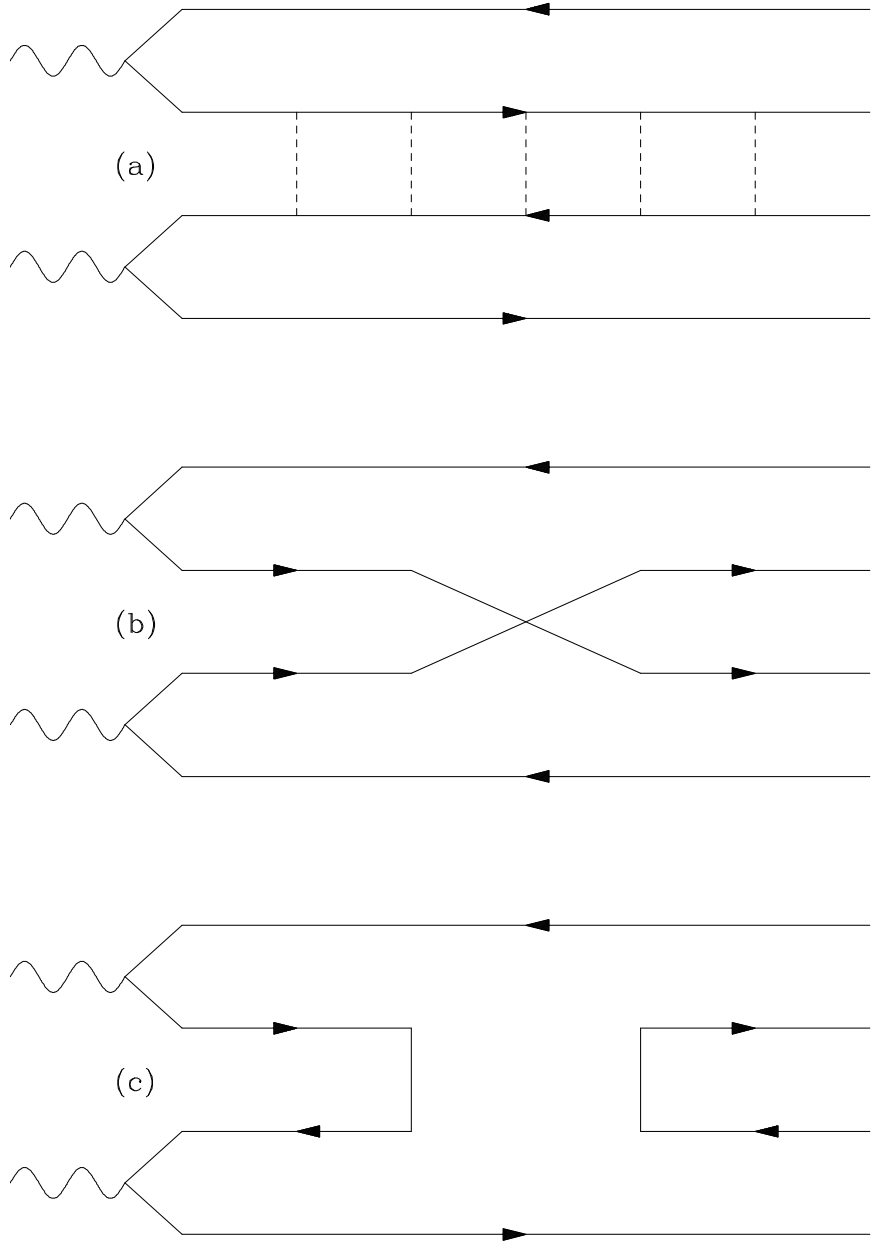}
\caption{Quark diagrams illustrating possible mechanisms for threshold
enhancement of vector meson pair production by two photons.  (a) Exchange
of $I=0$ mesons (dashed lines); (b) Quark exchange; (c) Quark-antiquark
annihilation and pair creation.}
\end{center}
\end{figure}

The presence of the threshold enhancement in $\gamma \gamma \to \rho^0 \rho^0$
in the state of total angular momentum $J=2$ and helicity $J_z = \pm 2$
\cite{Albrecht:1990cr} can be reproduced in resonance models
\cite{Achasov:1981kh,Li:1981ez} by suitable
choices of coupling constants.  It need not be universal.
In the model of Ref.\ \cite{Bajc:1996ew} the dominance of $(J,J_z) = (2,\pm 2)$ 
can be reproduced by spin-dependence in the potential between the vector
mesons.  If this potential is flavor-independent one would expect all possible
$\gamma \gamma \to V_1 V_2$ cross sections involving the diffractive process of
Fig.\ 1(a) to exhibit the same behavior as $\gamma \gamma \to \rho^0 \rho^0$:
a rapid rise near threshold in the $(J,J_z) =(2,\pm2)$ amplitude and little
activity in other partial waves.

Flavor-dependent effects which could enhance the $(J,J_z) =(2,\pm2)$ amplitude
include exchange of two identical quarks or quark-antiquark annihilation
between the two vector mesons, as illustrated in Fig.\ 1(b) and 1(c).
I assume that the initial vector mesons are produced only with $J_z = \pm 1$,
i.e., with the same polarizations as the incident photons.  Similar diagrams
were considered in a perturbative QCD calculation of non-diffractive $\gamma
\gamma \to V V$ processes \cite{Brodsky:1986ds}.

In the case of quark exchange, the quarks trading places in the vector mesons
have identical $J_z = \pm 1$ values only for the $J_z = \pm 2$ amplitudes,
while for the $J_z = 0$ amplitudes they necessarily have opposite $J_z$ values.
If the flavors of the two exchanged quarks do not matter, one can have
enhancements in processes $\gamma \gamma \to V_1 V_2$ where $V_1$ and $V_2$
are not able to couple directly to photons.  The initial process $\gamma \gamma
\to \rho^0 \rho^0$  could then lead to a $\rho^+ \rho^-$ final state, while
$\gamma \gamma \to \rho^0 \phi$ could lead to $K^{*+} K^{*-}$.  The absence of
such an enhancement in $\gamma \gamma \to \rho^+ \rho^-$ appears to disfavor
this mechanism.  One must then stipulate that the exchanged quarks be
identical, requiring additional dynamical assumptions.

In the case of quark-antiquark annihilation, as long as the annihilation must
take place in a state of $J_z = \pm 1$ (as, for example, in $q \bar q \to g$,
where $g$ is a transversely polarized gluon), the initial vector mesons must
have the same $J_z$ value.  The remaining $q' \bar q'$ pair will be left with
$J_z = \pm 1$.  If the transversely polarized gluon then materializes into a
different quark-antiquark pair, one will be left with a pair
of transversely polarized vector mesons different in flavor from the
previous one, so one should expect a threshold enhancement in $\gamma \gamma
\to \rho^+ \rho^-$, $\gamma \gamma \to K^{*+} K^{*-}$, and so on.  As in the
previous case, the absence of a threshold enhancement in $\gamma \gamma
\to \rho^+ \rho^-$ disfavors this alternative.

The relative strengths of the threshold enhancements in $\gamma \gamma \to
K^{*+} K^{*-}$ and $\gamma \gamma \to K^{*0} \bar K^{*0}$ do argue for some
contribution related to quark exchange and/or annihilation.  In either case
charged $K^*$ production is favored because one or both photons couples to
a $u$ quark with charge 2/3, while neutral $K^*$ production involves replacing
this quark with a $d$ quark with charge $-1/3$ \cite{Brodsky:1986ds}.
But an attempt to use the
diagrams of Fig.\ 1 for a unified description within flavor SU(3) founders
immediately on the inequality of the $\rho^+ \rho^-$ and $K^{*+} K^{*-}$ cross
sections, predicted to be equal within the U-spin subgroup of SU(3) involving
the interchange of $d$ and $s$ quarks. 

A further mechanism which could account for a threshold enhancement in $\gamma
\gamma \to \rho^0 \rho^0$ with $J_z = \pm 2$ is the effect of Bose statistics
in the final state \cite{Goldhaber:sf}.  A transversely polarized photon
with helicity $\lambda = \pm 1$ will produce a $\pi^+ \pi^-$ pair with
angular wave function proportional to $Y^1_\lambda(\theta,\phi)$, where
$(\theta,\phi)$ are the polar angles of the $\pi^+$ with respect to the
beam axis in the dipion rest frame.  At $\rho^0 \rho^0$ threshold the two
dipion rest frames coincide, leading to the possibility of coherent
reinforcement of processes in which both dipion systems have the same
wave functions $Y^1_\lambda(\theta_i,\phi_i)~(i=1,2)$.  As the center-of-mass
energy increases above threshold, the dipion wavefunctions refer to
individual rest frames which become distinct from one another, reducing the
possibility of coherence. 

The effects of Bose statistics (the ``GGLP effect'' \cite{Goldhaber:sf}) have
been observed in $\gamma \gamma \to 3 \pi^+ 3 \pi^-$ as an enchancement
in the low-$m(\pi^\pm \pi^\pm)$ distribution \cite{Pust:1991nj}.  However,
it appears that these effects have been ignored up to now in the
simpler process $\gamma \gamma \to 2 \pi^+ 2 \pi^-$.

The above realization of the GGLP mechanism implies threshold effects in
other cases of $\gamma \gamma \to V_1 V_2$ besides $V_1 = V_2 = \rho^0$, but
details may differ.  If the decay products of $V_1$ and $V_2$ are not all
identical, their overlaps will be reduced.  Such is the case, for example
in $\gamma \gamma \to \rho^+ \rho^- \to \pi^+ \pi^0 \pi^- \pi^0$, where only
the two neutral pions can overlap.  It is not even clear that under such
circumstances their relative phases would be the same as those of the
identical pions in $\gamma \gamma \to \rho^0 \rho^0 \to \pi^+ \pi^- \pi^+
\pi^-$.

If the decay product of the two vector mesons are identical, one is left
with the additional processes $\gamma \gamma \to \omega \omega,~\phi \phi,~
J/\psi J/\psi,~\ldots$, where both vector mesons decay to the same final
state.  Thus, in $\gamma \gamma \to \phi \phi$, a threshold enhancement should
be present when both $\phi$ mesons decay to $K^+ K^-$, but not when one decays
to $K^+ K^-$ and the other to $K_S K_L$.

The width of the decaying vector meson also probably plays a crucial role.
The probability of overlap of the decay products of two $\rho^0$ mesons is
very high since the $\rho^0$ has an appreciable width of about 150 MeV
\cite{PDG}.  Each $\rho^0$ thus can decay to $\pi^+ \pi^-$ in the vicinity
of the other $\rho^0$ before they move apart from one another.  The width
of the threshold enhancement (about 300 MeV) could then reflect the range
of center-of-mass energies over which this mechanism can occur.

The overlap of the decay products of two $\omega$ mesons is reduced for two
reasons.  First, the $\omega$ is much narrower, with a width of about 8 MeV
\cite{PDG}; second, the predominantly three-body decay of each $\omega$ implies
a much lower probability for overlap of the two vector mesons' decay products.

The process $\gamma \gamma \to \omega \omega$, having isospin $I=0$, cannot
benefit from $I=0$ -- $I=2$ interference \cite{Achasov:1981kh,Li:1981ez} to
account for the $\rho^0 \rho^0$ threshold enhancement.
It nonetheless exhibits a broad threshold enhancement \cite{Albrecht:1995nd},
rising to a maximum of about 4 nb near a center-of-mass energy of 2 GeV.
This is to be compared with the peak value of nearly 60 nb mentioned earlier
for $\gamma \gamma \to \rho^0 \rho^0$.  The reduced coupling of the photon to
the $\omega$ (1/3 of that to the $\rho$) more than explains this suppression;
in a naive vector-dominance model one would have expected 60/3$^4 \simeq
0.75$ nb.  Thus some mechanism in addition to the GGLP effect seems to be
in operation.

For $\gamma \gamma \to \phi \phi$ the $\sim 4$ MeV width of the $\phi$
\cite{PDG} implies that the decay products of each $\phi$ will be
well-separated from one another except just above threshold, so the GGLP
effect would imply {\it a very narrow threshold enhancement}, no more than a
few MeV wide, in $\gamma \gamma \to \phi \phi \to 2K^+ 2K^-$.
This enhancement should occur, as it does for $\gamma \gamma \to \rho^0
\rho^0$, in states with total two-photon helicity $J_Z = \pm 2$.  Up to now
there has been no reported observation of $\gamma \gamma \to \phi \phi$.

The GGLP effect discussed above has very different implications from some
other mechanisms for the threshold behavior of $\gamma \gamma \to V_1 V_2$,
where $V_1$ and $V_2$ are neutral mesons but $V_1 \ne V_2$.  For example,
the GGLP mechanism predicts no threshold enhancement in $\gamma \gamma
\to \pi^+ \pi^- K^+ K^-$ near $\rho^0 \phi$ threshold, whereas a
flavor-independent threshold attraction between vector mesons would imply
such an enhancement.

In fact, threshold enhancements are seen in {\it several} $\gamma \gamma \to
V_1 V_2~(V_1 \ne V_2)$ processes \cite{Kriznic:1994me}.  The cross section for
$\gamma \gamma \to \rho^0 \omega$ peaks at $17.3 \pm 3.1 \pm 1.7$ nb in the
center-of-mass energy range $1.5 < W < 1.7$ GeV.  The cross
section for $\gamma \gamma \to \rho^0 \phi$ peaks at $2.2 \pm 1.1 \pm 0.3$
nb in the range $1.75 < W < 2$ GeV.  A cross section
$\sigma(\gamma \gamma \to \omega \phi) = 1.65 \pm 0.86$ nb is measured
between 1.9 and 2.3 GeV.  All these cross sections fall off appreciably as
$W$ inceases.  The absence of interference between $V_1$ and
$V_2$ decay products makes it impossible to distinguish between $J_z = 0$
and $J_z = \pm 2$ production.

An approach which combines several aspects of the above proposals, known as
the ``threshold $t$-channel factorization model,'' is able to account for
some features of the data summarized in Table \ref{tab:pks}
\cite{Alexander:bv}.  It is applied primarily to diffractive processes for
which the diagram of Fig.\ 1(a) can contribute, with low-energy contributions
from other exchanges as well.  The combination of these effects can lead to
a peak near threshold.  The model has less to say about non-diffractive
processes in which only the flavor topologies of Figs.\ 1(b) and 1(c) are
relevant.

To summarize this section, there are appealing features of several of the
proposed models for threshold enhancements of $\gamma \gamma \to VV$, but
no obvious regularities in behavior that would permit one model to be
selected over others.  {\it Ad hoc} resonances, including exotic ones
\cite{Achasov:1981kh,Li:1981ez}, appear to require considerable fine-tuning if
they are to explain the observed pattern.  They nonetheless have the
considerable advantage of firmly predicting an $I=2$ resonance around 1600 MeV
which should decay to $\rho^\pm \rho^\pm$.  There are clearly some aspects of
double-diffractive production in the hierarchy $\sigma_{\rm pk}(\rho^0 \rho^0)
> \sigma_{\rm pk}(\rho^0 \omega) > \sigma_{\rm pk}(\omega \omega)$, reflecting
the hierarchy of photon--vector meson couplings, but SU(3) breaking is needed
to account for $\rho^0 \phi$ and $\omega \phi$ suppression, and evidence for
the expected hierarchy $\sigma_{\rm pk}(\rho^0 \phi) > \sigma_{\rm pk}(\omega
\phi)$ is fairly weak.  Altogether the situation calls for some new
insight and/or data.  The availability of photon-photon collisions
in larger samples amassed by the CLEO, BaBar, and Belle Collaborations
permits one not only to augment the statistics of the processes just mentioned,
but to investigate new ones, such as the process $\gamma \gamma \to J/\psi
\pi^+ \pi^-$ to which I now turn.
\bigskip

\centerline{\bf IV.  INFORMATION FROM $\gamma \gamma \to J/\psi \pi^+ \pi^-$}
\bigskip

The recently-studied process $\gamma \gamma \to J/\psi \pi^+ \pi^-$
\cite{Zweber} can provide further information on threshold enhancements in
vector meson pair production by two photons.  If a flavor-independent
interaction between vector mesons is responsible for enhanced production near
threshold, this process should exhibit such an effect.  The reaction $e^+ e^-
\to J/\psi \pi^+ \pi^- + X$ employed to study this process contains also
events in which one lepton loses a large amount of energy, $e^+ e^- \to \gamma
\psi' \to \gamma J/\psi \pi^+ \pi^-$, but backgrounds from these ``radiative
return'' events can be isolated by means of angular correlations of lepton
pairs in the $J/\psi$ decays.

If $\gamma \gamma \to J/\psi \pi^+ \pi^-$ is proceeding through a threshold
enhancement of $J/\psi \rho^0$ production related to the above effects,
several key features should be present in the data.  The vector mesons
should both be transversely polarized.  The leptons in $J/\psi \to \ell^+
\ell^-$ should then be distributed with probability $W_{\ell^+ \ell^-}
\sim 1 + \cos^2 \theta$ with respect to the beam axis, while the pions in
$\rho^0 \to \pi^+ \pi^-$ should have a distribution $W_{\pi^+ \pi^-} \sim
\sin^2 \theta$.  Pions arising from the radiative-return background $\psi'
\to \pi^+ \pi^- J/\psi$ should be emitted isotropically in their
center-of-mass system.  Unfortunately it will be impossible to distinguish
production with two-photon helicity $J_z = \pm 2$ from $J_z = 0$ since the
$J/\psi$ and $\rho^0$ decay products do not interfere with one another.

If a flavor-independent threshold enhancement mechanism is operative, the
cross section for $\gamma \gamma \to J/\psi \rho$ can be related to that for
(e.g.) $\gamma \gamma \to \phi \rho \sim 2$ nb by the scaling rule
\cite{Carlson:1974ev}
\begin{equation}
\frac{\sigma(\gamma \gamma \to J/\psi \rho)}
{\sigma(\gamma \gamma \to \phi \rho)} = \left( \frac{g_{J/\psi}}{g_\phi}
\right)^2 \left( \frac{M_\phi}{M_{J/\psi}} \right)^4
\left( \frac{\sigma(J/\psi \rho)}{\sigma(\phi \rho)} \right)^2~~,
\end{equation}
where $g_V$ is the coupling of vector meson $V$ to the photon, while
$\sigma(V_1 V_2)$ is the total cross section for scattering of vector mesons
$V_1$ and $V_2$ on each other.  Neglecting differences in masses and wave
functions, the couplings $g_V$ would scale as quark charges, entailing
$g_{J/\psi}/ g_\phi = -2$.  In fact the ratio of leptonic widths \cite{PDG}
implies
\begin{equation}
\left( \frac{g_{J/\psi}}{g_\phi} \right)^2 \left( \frac{M_\phi}{M_{J/\psi}}
\right)^3 = \frac{\Gamma_{ee}(J/\psi)}{\Gamma_{ee}(\phi)}
= \frac{(5.26 \pm 0.37)~{\rm keV}}{(1.26 \pm 0.02)~{\rm keV}} = 4.18 \pm 0.30~.
\end{equation}
The scaling arguments of Ref.\ \cite{Carlson:1974ev} imply that
\begin{equation}
\frac{\sigma(J/\psi \rho)}{\sigma(\phi \rho)} = \frac{M^2_{\phi}}{M^2_{J/\psi}}
\end{equation}
so that one predicts
\begin{equation} \label{eqn:Jpsirhopk}
\sigma(\gamma \gamma \to J/\psi \rho^0) = \sigma(\gamma \gamma \to \phi \rho^0)
\frac{M^5_{\phi}}{M^5_{J/\psi}} \frac{\Gamma_{ee}(J/\psi)}{\Gamma_{ee}(\phi)}
\simeq (26 \pm 13)~{\rm pb}~~.
\end{equation}
This estimate for the cross section near threshold is to be compared with a
range of 14--20 pb at $W=10$ GeV obtained in Ref.\ \cite{Goncalves:2003qq},
growing significantly at higher energies in a model-dependent manner.

The result (\ref{eqn:Jpsirhopk}) can be translated into a cross section for
$e^+ e^- \to e^+ e^- J/\psi \rho^0$ via two nearly real photons through the
relation
\cite{Brodsky:1970vk}
\begin{equation}
\sigma(e^+e^- \to e^+ e^- X) \simeq \left( \frac{2 \alpha}{\pi} \ln \frac{E}
{m_e} \right)^2 \int_{W_{th}}^{2E} \frac{dW}{W} f \left( \frac{W}{2E} \right)
\sigma_{\gamma \gamma \to X}(W)~~~,
\end{equation}
where $f(x) \equiv (2+x^2)^2 \ln (1/x) - (1-x^2)(3+x^2)$, $W$ is the
photon-photon center-of-mass energy, and $E$ is the electron or positron beam
energy (in a symmetric collider configuration).  The cross section
$\sigma_{\gamma \gamma \to X}(W)$ may be approximated by a Breit-Wigner form
\begin{equation}
\sigma_{\gamma \gamma \to X}(W) \simeq \frac{\sigma_{\gamma \gamma}^{\rm pk}}
{[2(W-W_0)/\Gamma]^2 + 1}~~~,
\end{equation}
where $W_0$ and $\Gamma$ are the mass and width of the resonance.  In the
narrow-resonance approximation (crude, but sufficient for our purposes) one
then has
\begin{equation} \label{eqn:ratio}
\sigma(e^+e^- \to e^+ e^- X) \simeq \frac{2 \Gamma}{\pi W_0} \left( \alpha
\ln \frac{E}{m_e} \right)^2 f \left( \frac{W_0}{2 E} \right)
\sigma_{\gamma \gamma}^{\rm pk}~~~.
\end{equation}
For $E = 5$ GeV, $\Gamma = 200$ MeV, $W_0 = 3.7$ GeV, one has $f(W_0/[2E])
= 1.83$, and Eq.\ (\ref{eqn:ratio}) gives $\sigma(e^+e^- \to e^+ e^- X)/
\sigma_{\gamma \gamma}^{\rm pk} \simeq 2.8 \times 10^{-4}$.  (Use of a slightly
more accurate expression \cite{Zweber} reduces this estimate by about 10\%.)
Thus, for a peak cross section given by Eq.\ (\ref{eqn:Jpsirhopk}), one
estimates $\sigma(e^+e^- \to e^+ e^- J/\psi \rho^0) \sim 7$ fb, with a 50\%
error.  Since the data sample reported in Ref.\ \cite{Zweber} consists of
15 fb$^{-1}$, one should see a handful of events in which the $J/\psi$ decays
to $\mu^+ \mu^-$ or $e^+ e^-$ in that sample, and considerably more in
samples accumulated by BaBar and Belle.

In calculating the sensitivity of a detector to $J/\psi \rho^0$ decays of a
resonance with definite spin $J$ and parity $P$ one needs the angular
distributions of final $\ell^+ \ell^- \pi^+ \pi^-$ systems associated with each
$J^P$ value.  Angular distributions have been treated in previous work (see,
e.g., Ref.\ \cite{Albrecht:1990cr}), but for $\ell^+ \ell^- \pi^+ \pi^-$ final
states great simplifications are possible using a {\it transversity basis}
\cite{Dunietz:1990cj,Dighe:1995pd}.  To illustrate the non-trivial nature of
these distributions it is helpful to compare them for the decays of $J^P = 0^+$
and $0^-$ particles into the lowest available partial waves, respectively S-
and P-waves.  One defines coordinate systems and three angles
$\psi,\theta,\varphi$ in the following manner \cite{Dighe:1995pd}.

In the rest frame of the $\pi^+ \pi^-$ system, the $x$ axis is defined as
the negative of the unit vector pointing in the direction of travel of the
$J/\psi$.  The $\pi^+ \pi^-$ system is assumed to lie in the $x$-$y$ plane,
with $\pi^+$ making an angle $\psi$ with the $x$ axis ($0 \le \psi \le \pi$).

The $z$ axis is taken in the $J/\psi$ rest frame perpendicular to the plane
containing the $\pi^+ \pi^-$ pair, using a right-handed coordinate system.  In
this frame the unit vector $\hat n(\ell^+)$ along the direction of the
positive lepton has coordinates $(n_x,n_y,n_z) = (\sin \theta \cos \varphi,
\sin \theta \sin \varphi, \cos \theta)$, thereby defining $\theta$ and $\phi$.

Applying the methods of Ref.\ \cite{Dighe:1995pd}, one then finds
\begin{equation}
\frac{1}{\Gamma} \frac{d^3 \Gamma}{d~\cos \theta d \varphi d~\cos \psi} =
\frac{9}{32 \pi} \left\{\begin{array}{l r} 1 - \sin^2 \theta \cos^2 (\psi -
\varphi) & (0^+) \cr \sin^2 \theta \sin^2 \psi & (0^-) \end{array} \right.
\end{equation}
for the differential angular distributions.  A clear difference is present.
The sensitivity of any detector to this difference will depend upon its
angular coverage and whether symmetric or asymmetric $e^+ e^-$ collisions are
employed.  With sufficient statistics, a distinction between even and odd
parity becomes possible.  Similar methods can be applied to the decays of
$2^{\pm}$ threshold enhancements.
\bigskip

\centerline{\bf V.  CONCLUSIONS}
\bigskip

The two-photon processes $\gamma \gamma \to V_1 V_2$, where $V_1$ and $V_2$
are vector mesons, display threshold enhancements in a variety of cases,
notably for $V_1 = V_2 = \rho^0$ but also elsewhere.  I have reviewed
several proposals for these enhancements and proposed others, such as
effects due to Bose statistics in decays of the vector mesons.  Tests
involving production of $\rho$, $\omega$, and $\phi$ mesons, performed over
many years, now can be augmented by the study of such processes as
$\gamma \gamma \to J/\psi \rho^0$, for which a simple model based on flavor
universality of the threshold enhancement predicts a cross section of
$26 \pm 13$ pb.  Angular distributions of decay products are illustrated for
spinless threshold enhancements and shown to be sensitive to parity.
\bigskip

\centerline{\bf ACKNOWLEDGMENTS}
\bigskip

I thank Roy Briere, Shmuel Nussinov, Kam Seth, David Urner, Helmut Vogel,
Jim Wiss, and Pete Zweber for discussions, Maury Tigner for extending the
hospitality of the Laboratory for Elementary-Particle Physics at Cornell during
this research, and Bernard Pire and Magno Machado for calling my attention to
Refs.\ \cite{Anikin:2003fr} and \cite{Goncalves:2003qq}.  This work
was supported in part by the United States Department of Energy through Grant
No.\ DE FG02 90ER40560 and in part by the John Simon Guggenheim Memorial
Foundation.

\def \ajp#1#2#3{Am.\ J. Phys.\ {\bf#1}, #2 (#3)}
\def \apny#1#2#3{Ann.\ Phys.\ (N.Y.) {\bf#1}, #2 (#3)}
\def \app#1#2#3{Acta Phys.\ Polonica {\bf#1}, #2 (#3)}
\def \arnps#1#2#3{Ann.\ Rev.\ Nucl.\ Part.\ Sci.\ {\bf#1}, #2 (#3)}
\def \art{and references therein}
\def \cmts#1#2#3{Comments on Nucl.\ Part.\ Phys.\ {\bf#1}, #2 (#3)}
\def \cn{Collaboration}
\def \cp89{{\it CP Violation,} edited by C. Jarlskog (World Scientific,
Singapore, 1989)}
\def \efi{Enrico Fermi Institute Report No.\ }
\def \epjc#1#2#3{Eur.\ Phys.\ J. C {\bf#1}, #2 (#3)}
\def \f79{{\it Proceedings of the 1979 International Symposium on Lepton and
Photon Interactions at High Energies,} Fermilab, August 23-29, 1979, ed. by
T. B. W. Kirk and H. D. I. Abarbanel (Fermi National Accelerator Laboratory,
Batavia, IL, 1979}
\def \hb87{{\it Proceeding of the 1987 International Symposium on Lepton and
Photon Interactions at High Energies,} Hamburg, 1987, ed. by W. Bartel
and R. R\"uckl (Nucl.\ Phys.\ B, Proc.\ Suppl., vol.\ 3) (North-Holland,
Amsterdam, 1988)}
\def \ib{{\it ibid.}~}
\def \ibj#1#2#3{~{\bf#1}, #2 (#3)}
\def \ichep72{{\it Proceedings of the XVI International Conference on High
Energy Physics}, Chicago and Batavia, Illinois, Sept. 6 -- 13, 1972,
edited by J. D. Jackson, A. Roberts, and R. Donaldson (Fermilab, Batavia,
IL, 1972)}
\def \ijmpa#1#2#3{Int.\ J.\ Mod.\ Phys.\ A {\bf#1}, #2 (#3)}
\def \ite{{\it et al.}}
\def \jhep#1#2#3{JHEP {\bf#1}, #2 (#3)}
\def \jpb#1#2#3{J.\ Phys.\ B {\bf#1}, #2 (#3)}
\def \lg{{\it Proceedings of the XIXth International Symposium on
Lepton and Photon Interactions,} Stanford, California, August 9--14 1999,
edited by J. Jaros and M. Peskin (World Scientific, Singapore, 2000)}
\def \lkl87{{\it Selected Topics in Electroweak Interactions} (Proceedings of
the Second Lake Louise Institute on New Frontiers in Particle Physics, 15 --
21 February, 1987), edited by J. M. Cameron \ite~(World Scientific, Singapore,
1987)}
\def \kdvs#1#2#3{{Kong.\ Danske Vid.\ Selsk., Matt-fys.\ Medd.} {\bf #1},
No.\ #2 (#3)}
\def \ky85{{\it Proceedings of the International Symposium on Lepton and
Photon Interactions at High Energy,} Kyoto, Aug.~19-24, 1985, edited by M.
Konuma and K. Takahashi (Kyoto Univ., Kyoto, 1985)}
\def \mpla#1#2#3{Mod.\ Phys.\ Lett.\ A {\bf#1}, #2 (#3)}
\def \nat#1#2#3{Nature {\bf#1}, #2 (#3)}
\def \nc#1#2#3{Nuovo Cim.\ {\bf#1}, #2 (#3)}
\def \nima#1#2#3{Nucl.\ Instr.\ Meth. A {\bf#1}, #2 (#3)}
\def \np#1#2#3{Nucl.\ Phys.\ {\bf#1}, #2 (#3)}
\def \npbps#1#2#3{Nucl.\ Phys.\ B Proc.\ Suppl.\ {\bf#1}, #2 (#3)}
\def \os{XXX International Conference on High Energy Physics, Osaka, Japan,
July 27 -- August 2, 2000}
\def \PDG{Particle Data Group, K. Hagiwara \ite, \prd{66}{010001}{2002}}
\def \pisma#1#2#3#4{Pis'ma Zh.\ Eksp.\ Teor.\ Fiz.\ {\bf#1}, #2 (#3) [JETP
Lett.\ {\bf#1}, #4 (#3)]}
\def \pl#1#2#3{Phys.\ Lett.\ {\bf#1}, #2 (#3)}
\def \pla#1#2#3{Phys.\ Lett.\ A {\bf#1}, #2 (#3)}
\def \plb#1#2#3{Phys.\ Lett.\ B {\bf#1}, #2 (#3)}
\def \pr#1#2#3{Phys.\ Rev.\ {\bf#1}, #2 (#3)}
\def \prc#1#2#3{Phys.\ Rev.\ C {\bf#1}, #2 (#3)}
\def \prd#1#2#3{Phys.\ Rev.\ D {\bf#1}, #2 (#3)}
\def \prl#1#2#3{Phys.\ Rev.\ Lett.\ {\bf#1}, #2 (#3)}
\def \prp#1#2#3{Phys.\ Rep.\ {\bf#1}, #2 (#3)}
\def \ptp#1#2#3{Prog.\ Theor.\ Phys.\ {\bf#1}, #2 (#3)}
\def \rmp#1#2#3{Rev.\ Mod.\ Phys.\ {\bf#1}, #2 (#3)}
\def \rp#1{~~~~~\ldots\ldots{\rm rp~}{#1}~~~~~}
\def \rpp#1#2#3{Rep.\ Prog.\ Phys.\ {\bf#1}, #2 (#3)}
\def \sing{{\it Proceedings of the 25th International Conference on High Energy
Physics, Singapore, Aug. 2--8, 1990}, edited by. K. K. Phua and Y. Yamaguchi
(Southeast Asia Physics Association, 1991)}
\def \slc87{{\it Proceedings of the Salt Lake City Meeting} (Division of
Particles and Fields, American Physical Society, Salt Lake City, Utah, 1987),
ed. by C. DeTar and J. S. Ball (World Scientific, Singapore, 1987)}
\def \slac89{{\it Proceedings of the XIVth International Symposium on
Lepton and Photon Interactions,} Stanford, California, 1989, edited by M.
Riordan (World Scientific, Singapore, 1990)}
\def \smass82{{\it Proceedings of the 1982 DPF Summer Study on Elementary
Particle Physics and Future Facilities}, Snowmass, Colorado, edited by R.
Donaldson, R. Gustafson, and F. Paige (World Scientific, Singapore, 1982)}
\def \smass90{{\it Research Directions for the Decade} (Proceedings of the
1990 Summer Study on High Energy Physics, June 25--July 13, Snowmass, Colorado),
edited by E. L. Berger (World Scientific, Singapore, 1992)}
\def \tasi{{\it Testing the Standard Model} (Proceedings of the 1990
Theoretical Advanced Study Institute in Elementary Particle Physics, Boulder,
Colorado, 3--27 June, 1990), edited by M. Cveti\v{c} and P. Langacker
(World Scientific, Singapore, 1991)}
\def \yaf#1#2#3#4{Yad.\ Fiz.\ {\bf#1}, #2 (#3) [Sov.\ J.\ Nucl.\ Phys.\
{\bf #1}, #4 (#3)]}
\def \zhetf#1#2#3#4#5#6{Zh.\ Eksp.\ Teor.\ Fiz.\ {\bf #1}, #2 (#3) [Sov.\
Phys.\ - JETP {\bf #4}, #5 (#6)]}
\def \zpc#1#2#3{Zeit.\ Phys.\ C {\bf#1}, #2 (#3)}
\def \zpd#1#2#3{Zeit.\ Phys.\ D {\bf#1}, #2 (#3)}

\end{document}